\documentclass[fleqn,10pt]{wlscirep}

\title{The anomalous tango of hemocyte migration in \emph{Drosophila melanogaster} embryos}

\author[1,*]{Nickolay Korabel}
\author[2]{Giuliana D. Clemente}
\author[1,3,4]{Daniel Han}
\author[4]{Felix Feldman}
\author[3,*]{Tom H. Millard}
\author[4,5,*]{Thomas Andrew Waigh}
\affil[1]{Department of Mathematics, The University of Manchester, M13 9PL, UK}
\affil[2]{School of Biochemistry, University of Bristol, BS8 1TD, UK}
\affil[3]{Faculty of Biology, Medicine and Health, The University of Manchester, M13 9PT, UK}
\affil[4]{Biological Physics, Department of Physics and Astronomy, The University of Manchester, M13 9PL, UK}
\affil[5]{Photon Science Institute, The University of Manchester, M13 9PL, UK}
\affil[*]{nickolay.korabel@manchester.ac.uk}
\affil[*]{tom.millard@manchester.ac.uk}
\affil[*]{t.a.waigh@manchester.ac.uk}
\affil {Nicolay Korabel and Giuliana Clemente contributed equally to the work}

\begin{abstract}
\emph{Drosophila melanogaster} hemocytes are highly motile cells that are crucial for successful embryogenesis and have important roles in the organism's immunological response. Hemocyte motion was measured using selective plane illumination microscopy. Every hemocyte cell in one half of an embryo was tracked during embryogenesis and analysed using a deep learning neural network. The anomalous transport of the cells was well described by fractional Brownian motion that was heterogeneous in both time and space. Hemocyte motion became less persistent over time. 
\emph{LanB1} and \emph{SCAR} mutants disrupted the collective cellular motion and reduced its persistence due to the modification of viscoelasticity and actin-based motility respectively. The anomalous motility of the hemocytes oscillated in time with alternating epoques of varying persistent motion. Touching hemocytes experience synchronised contact inhibition of locomotion; an anomalous tango. A quantitative statistical framework is presented for hemocyte motility which provides new biological insights.
\end{abstract}
\begin{document}

\flushbottom
\maketitle
\thispagestyle{empty}

\section*{Introduction}
Cellular motility is a crucial process in biology. It is essential for  morphogenesis, immune responses, cancer metastasis and wound healing. Although some progress has been made in understanding the motion of isolated eukaryotic cells on artificial substrates \cite{Metzner}, less is known about their collective behaviour \emph{in vivo}. A diverse selection of experimental results suggest that single cell movements can range from simple persistent random motion \cite{Selmeczi} and heterogeneous persistent random motion \cite{Wu2014,Wu2015} to Lévy-like movement patterns \cite{Huda} and Lévy walks \cite{Harris,Ariel}. Our quantitative understanding of cell motility inside organisms is much more patchy \cite{WirtzReview}. We are not aware of any previous detailed evidence for the anomalous diffusion of cells inside embryos. Single cell tracking has been previously performed in embryos (e.g. Zebra fish, \emph{Drosophila} and mice), but the resultant data was predominantly analyzed in terms of cell destinations to understand morphogenesis, combined with relatively qualitative measures of cell motility with small numbers of cells e.g. instantaneous velocities and apparent diffusion coefficients that do not accurately describe the signatures of anomalous transport of the whole ensemble \cite{Chen,Tomer,Schott}. Some work has been performed on glassy dynamics in jammed tissue from zebra fish embryos \cite{Kim,Schotz,Goodwin}, but this is predominantly from a mechanical perspective. Here we focus on cell motility in \emph{Drosophila} embryos using advanced optical microscopy techniques.

A major barrier in cell motility research has been experimental, since cellular aggregates are fragile, often opaque and it is challenging to image them \emph{in vivo} with sufficient resolution without causing damage. We used selective plane illumination microscopy (SPIM) to image the migration of hemocyte cells in developing \emph{Drosophila melanogaster} embryos during early stages of their morphogenesis \cite{Huisken}. The SPIM experiments allowed hemocytes to be non-invasively imaged for 6 hours. The nuclei of hemocytes were made fluorescent using genetically encoded histone tags and this allowed the motility of every hemocyte to be imaged and subsequently tracked in three dimensions across half an embryo. This would not be possible with conventional confocal microscopy techniques due to radiation damage and has not before been achieved with hemocytes. Previous live analyses of hemocyte migration have focused on relatively few cells migrating in a small and defined region of the embryo, most frequently the narrow space between the ventral nerve cord and epidermis. Previously we developed a deep learning neural network analysis tool to analyze two dimensional tracks of intracellular endosomes that allowed automatic dynamic segmentation of the tracks in terms of fractional Brownian motion (FBM) with anomalous diffusion exponents and generalized diffusion coefficients that are heterogeneous in both space and time \cite{Korabel_etal,Korabel_etal_Entropy}. A new theory for the heterogeneous anomalous transport was developed which was made possible by the order of magnitude improvement in sensitivity provided by the neural network \cite{han2020deciphering}. In the current study, we extend our neural network methodology for anomalous transport to the motion of hemocyte cells in three dimensions inside embryos and present a new model for their motility based on new SPIM data.

It has been observed in many experiments that due to viscoelasticity and heterogeneity of the environment, obstructions, spatio-temporal correlations and non-specific interactions; intracellular transport and cell motility do not follow Brownian diffusion \cite{Caspi,Granick1,Granick2,PLOS,Fedotov,Kenwright,Cherstvy} or simple persistent random walks \cite{Wu2014,Wu2015}. 
Single cell tracking of {\it Dictyostelium discoideum} cells revealed
anomalous diffusion and power-law tails for the velocity distributions which were explained by a generalized Langevin model \cite{Takagi}. Anomalous diffusion was also found in 
trajectories of single wild-type and mutated epithelial (transformed
Madin–Darby canine kidney) cells \cite{Klages}. The super-diffusive increase of the mean squared displacement, non-Gaussian displacement probability distributions, and power-law decays of the velocity autocorrelations were interpreted using a fractional Klein–Kramers equation. Single metastatic cancer cells migrating on linear microtracks were shown to follow Lévy-like patterns of movement \cite{Huda}. Even more complex motility was observed  in tracks of single CD8+ T cell migration, which were found to be consistent with generalized Lévy walks \cite{Harris}. The simultaneous motility of  multiple cells is expected to impact the forms of anomalous transport demonstrated by single cells e.g. via jamming, contact inhibition, intercellular signalling and complex three dimensional geometries \cite{Alert, Romanczuk}. We therefore made a quantitative study of the anomalous transport of hemocyte cells in \emph{Drosophila} embryos.  

There are four phases of development for \emph{Drosophila melanogaster}. The embryonic phase is the first of these four stages and it is the focus of the current study. During embryogenesis, hemocytes are formed at a single site in the developing head and undergo a process called dispersal in which they migrate away from their origin over the course of several hours until they are spread throughout the embryo. The hemocytes are involved in phagocytosis (immune responses), encapsulation, melanization and extracellular matrix production \cite{Wood}. In general, cell migration can be either passive or active. Cells can be passively carried by fluid flow of the circulatory system or the movement of the surrounding tissue e.g. during germ band retraction. Active migration involves a crawling motion of the cells, driven by self-assembly of cytoskeletal proteins inside the cells. The motility is modulated by adhesion with the surrounding cells and the extracellular matrix (ECM), viscoelasticity and confinement effects (there are well defined pathways through the embryo and haemocytes experience contact inhibition) \cite{Wood}. In addition the cells' motion is directed by extracellular signals \cite{Wood}. These can attract the cells to migrate along specific routes and cause them to become oriented. Chemoattractants for hemocyte motion include Pvf (important for guiding migration) and factors involved in contact inhibition of locomotion (CIL) in which hemocyte cells repel one another \cite{Pocha, Davis, Wood2006}. Pvf and CIL are thought to be the main mechanisms that guide hemocyte migration during dispersal.

In the current study, two separate mutations were examined in addition to wild type \emph{Drosophila}, called \emph{LanB1} and \emph{SCAR}. \emph{LanB1} has the beta subunit of laminin deleted, which affects the extracellular matrix (ECM) and thus the localized viscoelasticity experienced by the hemocytes, which in turn affects their migration \cite{Sanchez}. In \emph{SCAR} mutant embryos, SCAR, a protein that promotes nucleation of actin filaments at the leading edge of the migrating cells in absent \cite{Evans}. Thus it was possible to separate the effects of ECM viscoelasticity and internal cellular dynamics on the  
 anomalous transport of the hemocytes.

\section*{Results}

\subsection*{Quantitative characterisation of temporal heterogeneity in hemocyte migration}

\begin{figure}
    \centering
    \includegraphics[width=\linewidth]{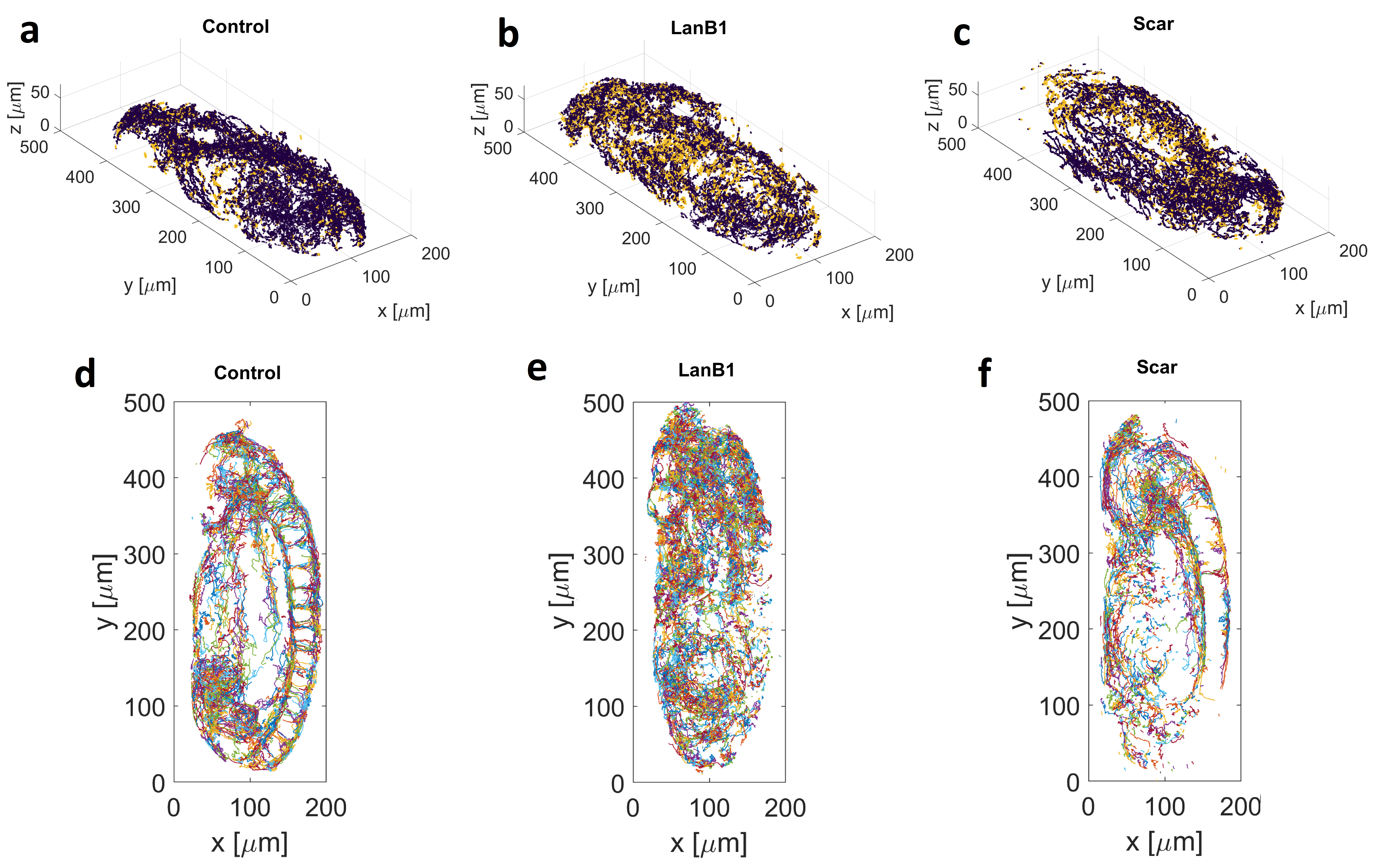}
    \caption{{\bf Hemocytes randomly transition between persistent and anti-persistent motion in different proportions while exploring control, \emph{LanB1} and \emph{SCAR} embryos. (a, b, c)} Three dimensional hemocyte trajectories with colours corresponding to local persistent ($H(t)>0.5$)  (violet) and anti-persistent ($H(t)<0.5$) (yellow) motion. Here $H(t)$ is the local Hurst exponent estimated using the neural network. Trajectories in \emph{SCAR} experiment have a greater amount of anti-persistent motion compared to the control and {\it LanB1}. {\bf (d, e, f)} Two dimensional projections of all the tracks reveal large scale dynamical patterns and anatomical features of the embryos become visible.
    }
    \label{fig:3dtracks}
\end{figure}

We quantified the heterogeneity of hemocyte motion via the analysis of their local dynamics in wild-type embryos and embryos with loss of function mutations in genes encoding \emph{SCAR} and \emph{LanB1}. We used a neural network method \cite{han2020deciphering} and estimated the local Hurst exponents $H(t)$ of individual hemocyte trajectories. The spatio-temporal dynamics were then classified as persistent $H(t)>0.5$ or anti-persistent $H(t)<0.5$. Trajectories in three dimensions displayed different proportions of persistent and anti-persistent movement for the control, {\it LanB1} and {\it SCAR} embryos (Fig.\ \ref{fig:3dtracks}). All the instantaneous values of the Hurst exponent can be plotted together for an embryo and anatomical details of the \emph{Drosophila} embryos are observed. The instantaneous Hurst exponents illustrate distinct changes in their spatial arrangements for the control and the two mutants ({\it LanB1} and {\it SCAR}) (Fig.\ \ref{fig:3dtracks}). Trajectories in {\it LanB1} and {\it SCAR} experiments display progressively greater amounts of anti-persistent motion compared to the control. 

\subsection*{Heterogeneous anomalous diffusion of hemocytes: non-Gaussian displacement distributions, super-diffusive MSDs and power-law decaying VACFs}

\begin{figure}
    \centering
    \includegraphics[width=\linewidth]{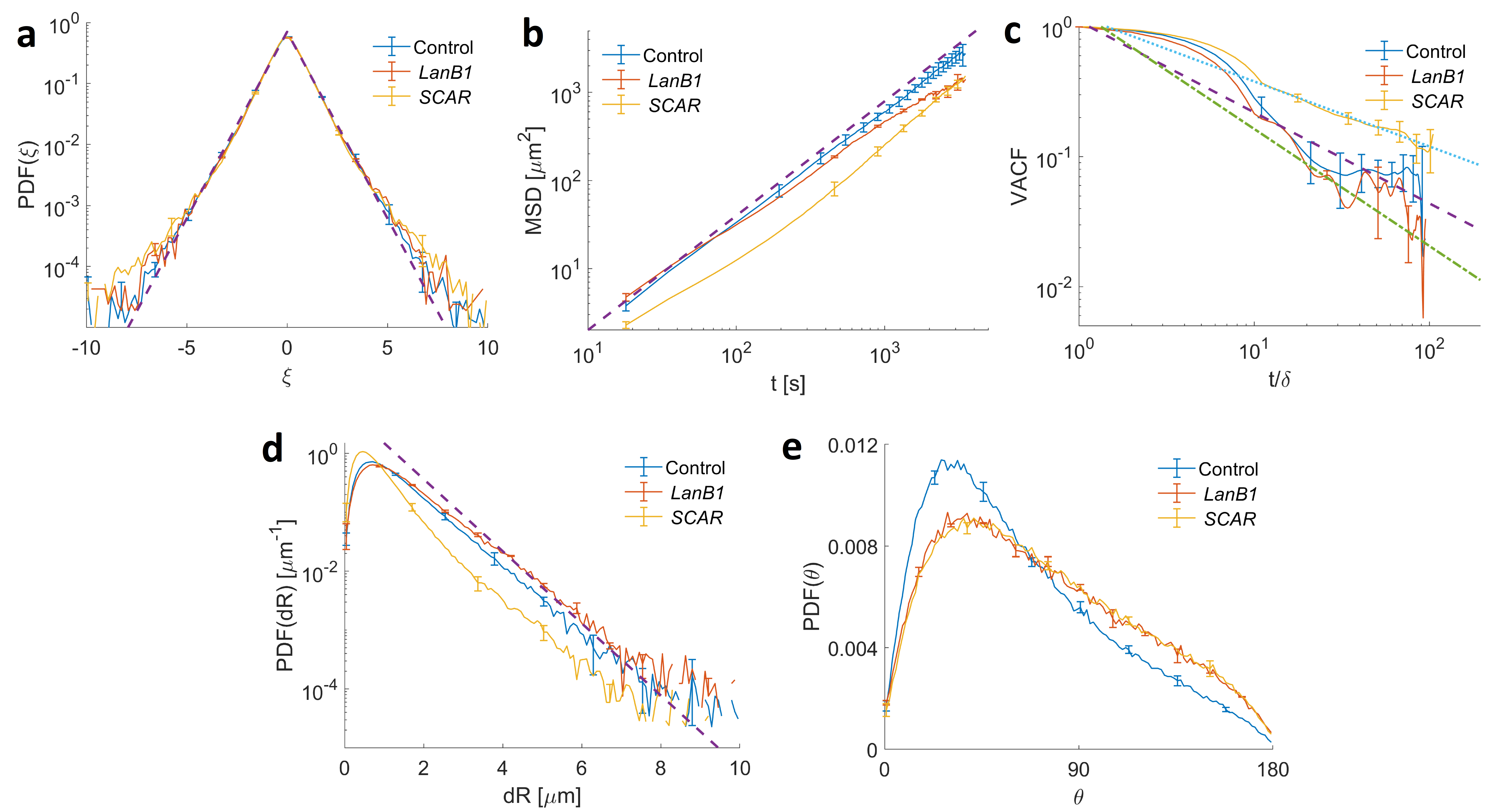}
    \caption{{\bf Hemocytes in control, \emph{LanB1} and \emph{SCAR} embryos move via heterogeneous anomalous diffusion. (a)} Distributions of displacements, $dx=x(t+\tau)-x(t)$, calculated at $\tau=17.6$ s are non-Gaussian and follow the Laplace distribution (the dashed line), $PDF(\xi)=\exp \left( - |\xi|/b \right)/2b$ with $\xi=dx/\sigma_x$ and $b=0.71$. {\bf (b)}  MSDs and TMSDs as a function of time interval grow superdiffusively with anomalous exponents $\alpha=1.3 \pm 0.1$ in the  control and $\alpha=1.1 \pm 0.1$ for {\it LanB1} embryo. {\it LanB1} and \emph{SCAR} embryos have progressively more anti-persistent motion on the shortest time scales while their long time behaviour remains superdiffusive. At longer time scales the MSD of Scar increases with anomalous exponents $\alpha=1.5 \pm 0.1$ (see Fig. \ref{figS1}). The dashed line represents the power-law functions $t^{1.3}$ for the control as a guide. {\bf (c)} Velocity auto-correlation functions as a function of time interval are positive and decay to zero as power-laws, $(t/\delta)^{\alpha-2}$, with $\alpha$ determined from the MSDs. {\bf (d)} Distributions of incremental displacements $dR = \sqrt{dx^2 + dy^2 + dz^2}$ calculated for 17.6 s time intervals decay exponentially. The dashed line represents an exponential function (the dashed line). {\bf (e)} PDFs of incremental angles $\theta$ for 17.6 time intervals calculated using the dot product of consecutive position vectors, $\bf{r_1} = (x_1,y_1,z_1)$ and $\bf{r_2} = (x_2,y_2,z_2)$, as 
    $\theta = \arccos(\bf{r_1}\cdot \bf{r_2}/|\bf{r_1}||\bf{r_2}|)$. Statistical quantities in this figure were averaged over $5$ data sets of control, $2$ {\it LanB1}, $4$ {\it SCAR} embryos. The error bars correspond to standard error of the mean.
    }
    \label{fig2}
\end{figure}

The probability distributions of hemocyte displacements scaled by their standard deviation were plotted at the shortest time interval probed (Fig. 2(a)). The distributions are clearly non-Gaussian and possess exponential tails (Laplace-like distributions). Over the range of time scales probed, the rescaled distributions for the control, {\it LanB1} and \emph{SCAR} are almost indistinguishable. Fig. \ref{fig2}(b) shows the mean square displacement (MSD) and time averaged mean squared displacement (TMSD) of the cells as a function of time interval. The control, {\it LanB1} and \emph{SCAR} MSDs are all superdiffusive ($\alpha>1$) and the amplitude of the motion decreases in the order control$>${\it LanB1}$>$\emph{SCAR}. Thus the mutations are not noticeably affecting the form of the statistics of the displacement distribution of the motility (Fig.\ \ref{fig2}(a)), rather it is the changes in displacement magnitude (Fig.\ \ref{fig2}(b)) and direction (persistence, Fig. 1) that are having a major impact. The velocity autocorrelation functions (VACF) as a function of time interval (Fig.\ \ref{fig2}(c)) show similar decays for the control and {\it LanB1} sample, whereas the \emph{SCAR} sample is much slower to decay.

The ensemble averaged mean squared displacements (MSD) (Fig.\ \ref{fig2}(b)) demonstrate a clear super-diffusive growth ($\alpha>1$) in the control, \emph{LanB1} and \emph{SCAR} embryos, indicating actively driven motion. Ensemble MSDs and time averaged TMSDs have near identical behaviour (see Fig. \ref{figS1}) which suggests ergodic motility for the hemocytes \cite{Metzler2014}. At short time scales, the slope of the MSDs is smaller for {\it LanB1} and \emph{SCAR} in agreement with the increased amount of anti-persistent motion observed in Fig.\ \ref{fig:3dtracks} (a,b,c). The velocity auto-correlation functions (VACF) (Fig.\ \ref{fig2}(c)) are positive for all the embryos studied and decay to zero as power-laws, $t^{\alpha-2}$, with $\alpha$ in agreement with that from the MSDs. The power-law behaviour of the MSDs and VACFs together with ergodic behaviour of the MSDs (Fig.\ \ref{fig2}(b,c)) suggest that hemocytes are described by fractional Brownian motion (FBM) or the fractional Langevin equation, rather than non-ergodic continuous time random walks (CTRW). These three models (FBM, FLE and CTRW) are the main possibilities used to describe anomalous transport in the literature i.e. to motivate fractional values of $\alpha$. This supports our decision to train the neural network using FBM trajectories \cite{han2020deciphering}. The distributions of incremental displacements $dR=\sqrt{dx^2 + dy^2 + dz^2}$ with $dx=x(t+\tau)-x(t)$, $dy=y(t+\tau)-y(t)$ and $dz=z(t+\tau)-z(t)$ also decay exponentially (Fig.\ \ref{fig2}(d)) similar to the distributions of displacements (Fig.\ \ref{fig2}(a)). However, they show that hemocytes in the \emph{SCAR} embryos are characterized by smaller incremental displacements compared to the  control and {\it LanB1} mutants. From Fig. \ref{fig2}(e) we also calculated the mean cell speed $\bar{v}=0.066\pm 0.046$ $\mu$m/s for control, $\bar{v}=0.074\pm 0.05$ $\mu$m/s for \emph{LanB1} and $\bar{v}=0.045\pm 0.033$ $\mu$m/s for \emph{SCAR}. Hemocytes in control embryos move persistently compared to \emph{LanB1} and \emph{SCAR} embryos. Corresponding distributions of incremental angles are shown in Fig.\ \ref{fig2}(e). This is consistent with our observation that the amount of persistent motion decreases in the order control$>${\it LanB1}$>$\emph{SCAR} (Fig.\ \ref{fig2}(a,b,c)).
The non-Gaussian form of the distributions of the displacements (Fig.\ \ref{fig2}(a)) suggests that the FBM-like motility of hemocytes is heterogeneous. Below we explore the heterogeneity of hemocyte motion in detail by studying their local dynamics.

\subsection*{Distributions of hemocytes' local anomalous exponents and local generalized diffusion coefficients}

\begin{figure}
    \centering
    \includegraphics[width=\linewidth]{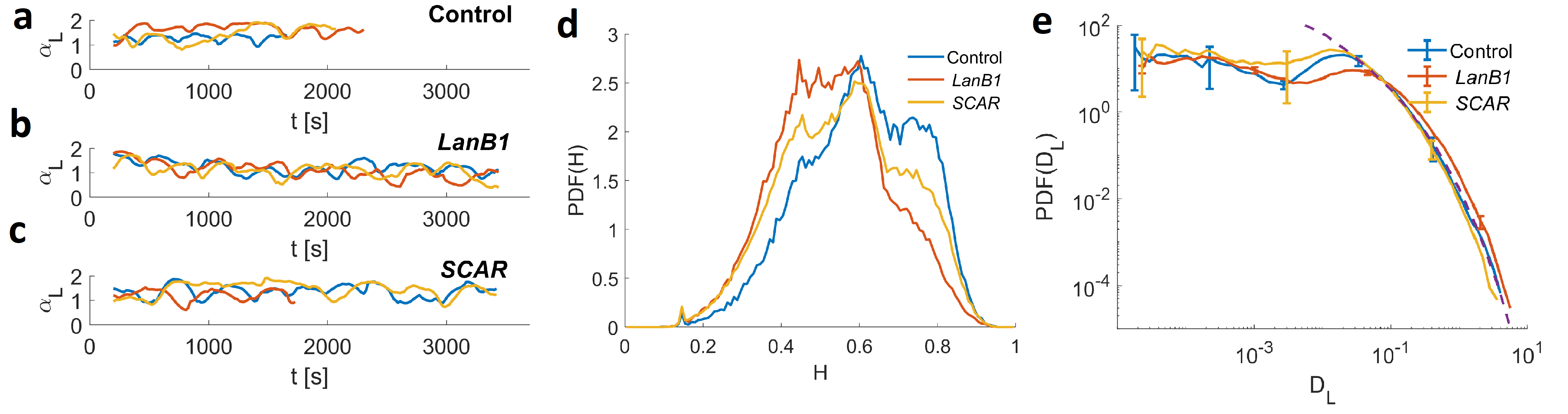}
    \caption{{\bf Local characterization of hemocytes' heterogeneous anomalous diffusion in control, \emph{LanB1} and \emph{SCAR} embryos. (a), (b), (c)} Local anomalous exponents $\alpha_L(t)$ for representative hemocyte trajectories in a control, {\it LanB1} and {\it SCAR} embryo. $\alpha_L(t)$ display oscillatory behaviour. 
    {\bf (d)} Distributions of time dependent Hurst exponents $H(t)$ estimated using the neural network. {\bf (e)} Distributions of local generalized diffusion coefficients $D_{\alpha_L}$. The dashed line represents a fit of the distribution tail with the Weibull density function, $PDF(\eta) \sim c (\eta^{c-1}/\eta_0^{c}) \exp(-(\eta/\eta_0)^c)$ with $c=0.39$ and $\eta_0=0.008$. The distributions were averaged over $5$ data sets of control, $2$ {\it LanB1}, $4$ {\it SCAR} embryos. The error bars correspond to standard error of the mean.
    }
    \label{fig3}
\end{figure}

To study the heterogeneity of hemocyte motion, we first calculated 
time dependent Hurst exponents $H(t)$ for individual hemocyte trajectories using the neural network \cite{han2020deciphering} for the control, {\it LanB1} and \emph{SCAR} embryos. For individual trajectories we also calculated local time averaged MSDs and extracted time dependent local anomalous exponents $\alpha_L(t)$ and local generalized diffusion coefficients $D_{\alpha_L}(t)$. Time dependent Hurst exponents $H(t)$ and time dependent local anomalous exponents $\alpha_L(t)$ quantify similar information (for standard FBM $\alpha = 2 H$), but $H(t)$ is determined more accurately by the neural network \cite{han2020deciphering}. We did both methods of analysis to provide an accurate test of the neural network analysis, bench-marked against a more traditional methodology. The typical behaviour of $\alpha_L(t)$ and $D_{\alpha_L}(t)$ calculated for sample trajectories in the {\it LanB1} embryo (Fig.\ \ref{fig3}(a,b)) reveals that $\alpha_L(t)$ and $D_{\alpha_L}(t)$ display oscillatory behaviour and are anti-correlated. A Fast Fourier Transform of $\alpha_L(t)$ (Fig.\ \ref{fig3}(c)) (should be SI Fig.3) reveals two periods of oscillation of approximately $3$ and $8$ minutes for the control, which vary for {\it LanB1} and {\it SCAR}. Time dependent Hurst exponents $H(t)$ for the control, {\it LanB1} and \emph{SCAR} embryos have broad distributions with a decreasing amount of persistent motion, following control$>${\it LanB1}$>$\emph{SCAR} and an increasing amount of  anti-persistent movement with control$<${\it LanB1}$<$\emph{SCAR} (Fig.\ \ref{fig3}(d)). The distributions of local generalized diffusion coefficients $D_{\alpha_L}(t)$ were found to decay with a Weibull distribution form (Fig.\ \ref{fig3}(e)) and \emph{SCAR} showed large numbers of small values of $D_{\alpha_L}(t)$.

To quantify the changes in proportion of persistent and anti-persistent motion, we used Gaussian mixture models to fit the distributions of local Hurst exponents $H(t)$. Four separate Gaussian modes were chosen to describe the data whose number was determined from the BIC criteria (SI, Fig. \ref{fig:gmmh}). The individual Gaussian components in the mixture models show a significant difference in modes of persistent ($H(t)>0.5$) motion between the control, \emph{SCAR} and {\it LanB1} embryos. The Hurst exponents of modes 1, 2 and 4 decrease with time (they become less persistent), whereas 3 is relatively constant.

\begin{figure}
    \centering
    \includegraphics[width=0.4\textwidth]{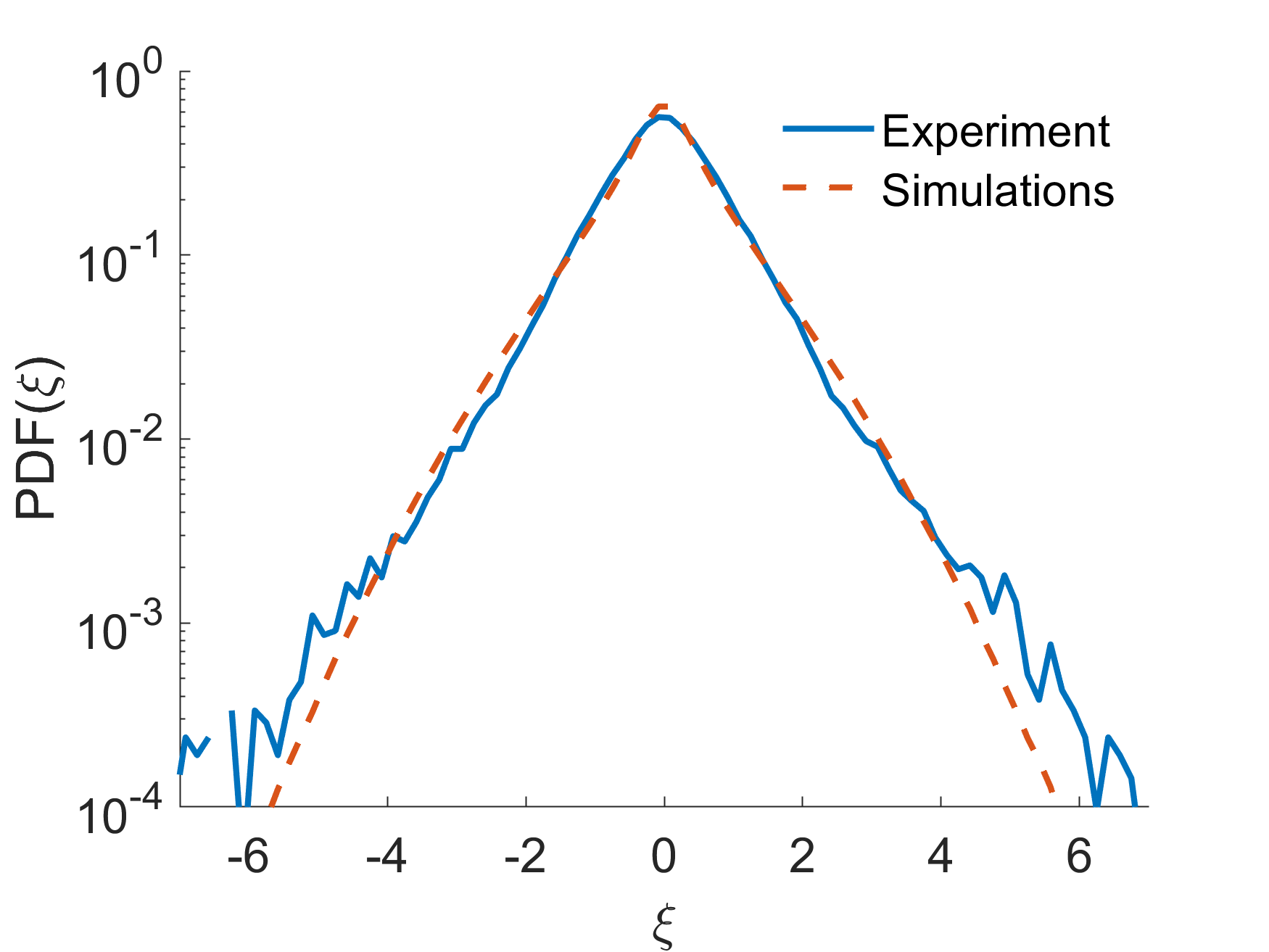}
    \caption{{\bf Heterogeneous FBM (hFBM) with the Weibull distribution of generalized diffusion coefficients reproduces displacements distributions in control, \emph{LanB1} and \emph{SCAR} embryos.} For clarity only the distribution of displacements in {\it LanB1} is shown. 
    }
    \label{fig4}
\end{figure}

The distributions of generalized diffusion coefficients $D_{\alpha_L}(t)$ obtained by calculating the local MSDs of individual hemocyte trajectories and fitting them with power-law functions, have Weibull distribution tails (Fig.\ \ref{fig3}(e)) for the control, {\it LanB1} and \emph{SCAR} embryos. Numerical simulations of heterogeneous FBM with a Weibull distribution of generalized diffusion coefficients (Fig.\ \ref{fig4}) are in good agreement with the Laplace form of the displacements  observed in the control, {\it LanB1} and \emph{SCAR} embryos (Fig.\ \ref{fig2}(a)).
 
 \subsection*{Oscillatory movement of hemocytes}

Strikingly, we found that the persistence of the motion of single cells is oscillatory with reasonably well defined time periods of $3 \pm0.5$ and $8 \pm0.5$ minutes for the wild type and $8 \pm0.5$ and $14 \pm0.5$ minutes (with small remnants of the 3 minute periodicity) for \emph{LanB1} and \emph{SCAR} (Fig.\ \ref{fig3}(a,b,c)). This is three orders of magnitude slower that the heart beat of \emph{Drosophila melanogaster} which is in the range 4-6 beats/s and the heart does not start to beat until later stages of development. The motion of the control, \emph{LanB1} and \emph{SCAR} were all found to be oscillatory. The absolute values of the anomalous exponents of single cell motility varied in space and time, which could be driven by varying viscoelasticities, confinement, adhesion, jamming or intracellular signalling (probably a combination of all these factors). It is expected the time period is internally driven by the cell, whereas the absolute values of the persistence of the motion are modulated by the external environment.

\begin{figure}
    \centering
    \includegraphics[width=0.7\textwidth]{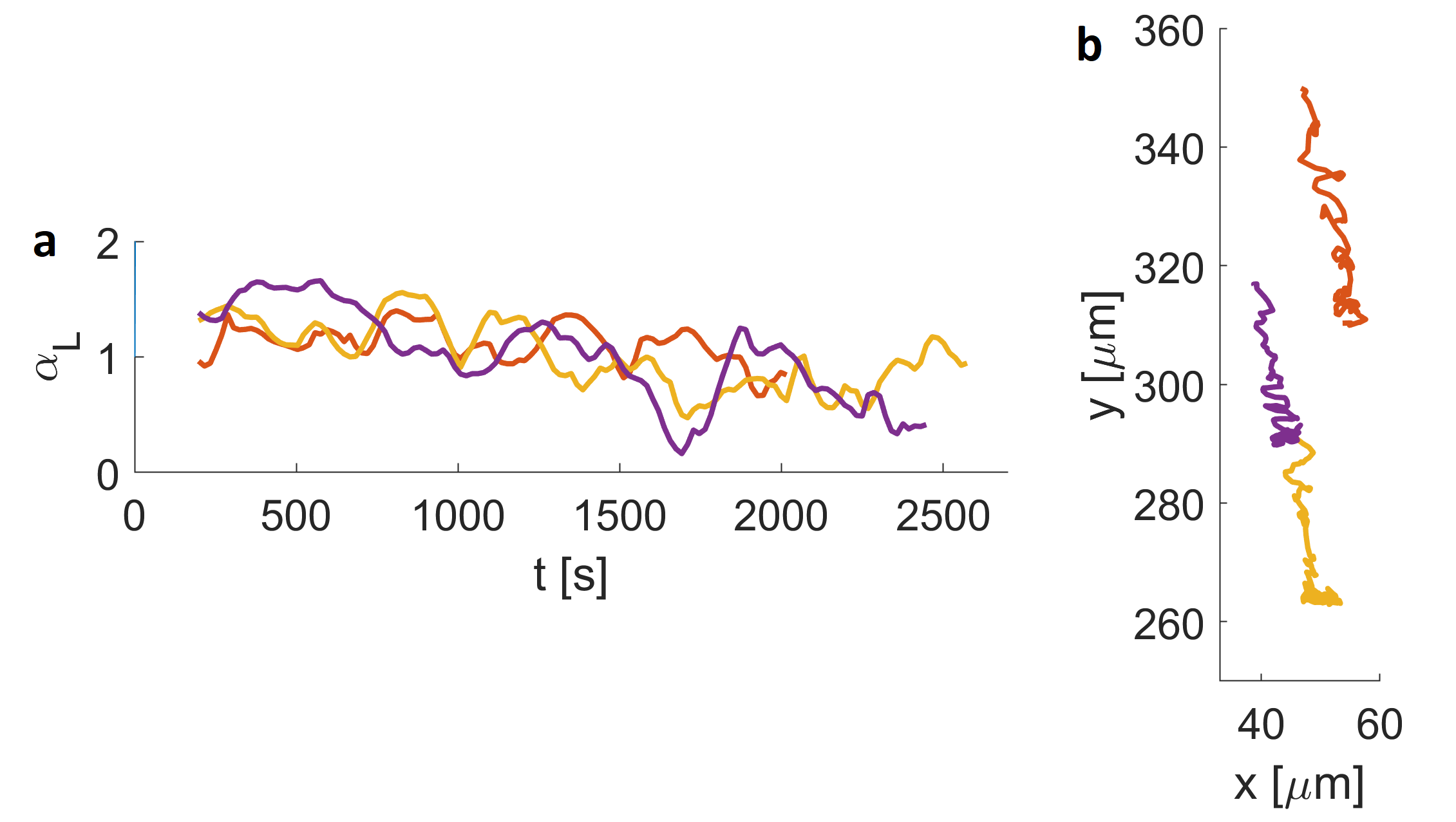}
    \caption{{\bf Coherent oscillations of local anomalous exponents reveal synchronised movement of hemocytes. (a)} Oscillating local anomalous exponents of sample trajectories of hemocytes in {\it LanB1}. {\bf (b)} Corresponding trajectories of hemocytes which are moving in a synchronised manner. We quantified correlations between trajectories (see SM Fig.\ \ref{fig:oclAlf}). Pairs of trajectories in {\bf (b)} have correlation coefficients close to $1$.
    }
    \label{fig5}
\end{figure}

The oscillatory behaviour of local anomalous exponents was synchronised for the local motion of non-contacting hemocytes. We found several nearby trajectories (Fig.\ \ref{fig5}(b)) which are similar in space and their local anomalous exponents display coherent oscillations in time (Fig.\ \ref{fig5}(a)). The slight time delays between the oscillations allow us to discount extrinsic experimental factors, such as instrument vibrations (the displacement amplitudes are also much larger than the static error in the tracking analysis). Furthermore, the neighbouring trajectories in Fig.\ \ref{fig5}(b) have displacement correlation coefficients approaching 1, whereas more distant cells are found to have much lower values e.g. some are negatively correlated (SI Fig.\ \ref{fig:oclAlf}). This suggests that local anomalous exponents are a valuable tool to detect synchronised motion of migrating cells in embryos. 

\begin{figure}
    \centering
    \includegraphics[width=0.9\textwidth]{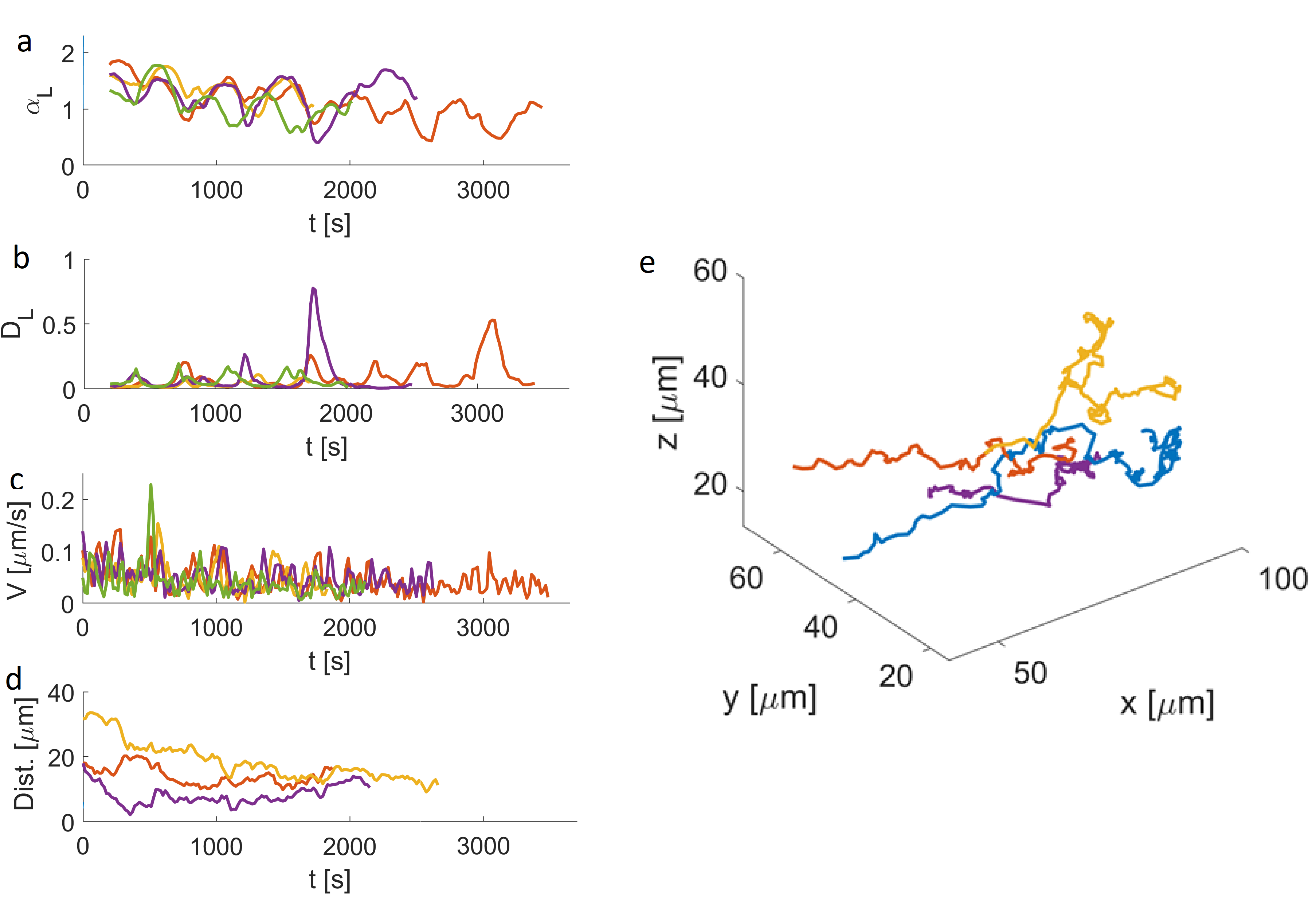}
    \caption{{\bf Contact inhibition leads to synchronization of hemocyte movement.} Oscillations of local anomalous exponents {\bf (a)}, local generalized diffusion coefficients {\bf (b)} and velocity {\bf (c)} synchronize when hemocytes experience contact inhibition. Hemocytes during the contact remain in close proximity to each other performing a tango of anomalous diffusion. {\bf (d)} Shows the relative distance between contacted hemocytes depicted in {\bf (e)}. Correlation coefficients for trajectories shown in {\bf (e)} calculated by Corr$(\lVert \bf{r_i} \rVert, \lVert \bf{r_j} \rVert)$ where $\lVert \bf{r_i} \rVert$ and $\lVert \bf{r_j} \rVert$ are distances vectors of trajectories $i$ and $j$. The correlation coefficients of three pairs of trajectories were $-0.08$, $0.52$ and $0.86$ indicating that trajectories remain correlated.
    }
    \label{fig6}
\end{figure}

Upon their contact, the hemocyte motion was also synchronised, although their clear sense of directional motion was lost (Fig.\ \ref{fig6}). Contact of the cells inhibits their motion (a process of contact inhibitory locomotion \cite{Stramer}) and their anomalous transport continues to be synchronised in terms of the $\alpha_L(t)$, $D_{\alpha_L}(t)$, displacements and the velocities (Fig.\ \ref{fig6} a,b, c and d). An anomalous tango is thus observed between the cells when they touch (Fig.\ \ref{fig6} e).

\subsection*{Truncated power-law survival times in persistent and anti-persistent hemocyte migration}

\begin{figure}
    \centering
    \includegraphics[width=0.5\textwidth]{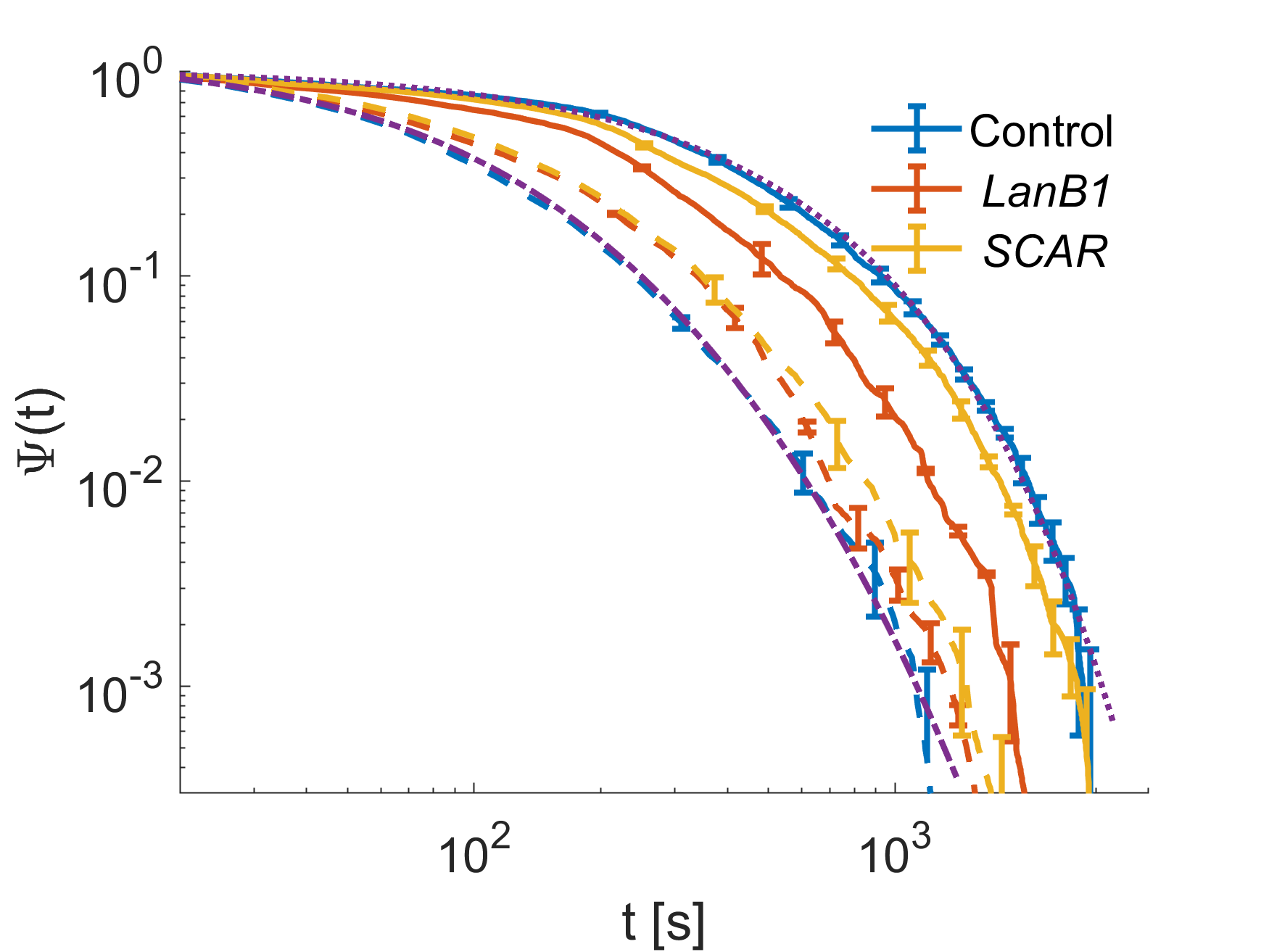}
    \caption{{\bf Survival functions of the times hemocytes move in persistent and anti-persistent motion in the control, {\it LanB1} and {\it SCAR} embryos are truncated power-laws.} Survival functions $\Psi(t)$ of persistent motion (solid curves) and anti-persistent motion (dashed curves) for control (blue), {\it LanB1} (red) and {\it SCAR} (yellow) embryos. Survival functions were averaged over $5$ data sets of control, $2$ {\it LanB1}, $4$ {\it SCAR} embryos. The error bars correspond to standard error of the mean. The dashed-doted and dotted curves are truncated power-law fits given by $(a/(a+t))^{b} e^{-c t}$  with $a=300$, $b=3.3$, $c=0.0016$ and $a=405$, $b=0.33$, $c=0.002$, respectively.
    }
    \label{fig7}
\end{figure}

Hemocytes constantly switch between persistent and anti-persistent motion (Fig.\ \ref{fig:3dtracks}). To quantify the amount of persistent and anti-persistent motion in the control and the two mutant embryos, we calculated the survival  functions of the times the hemocytes spent in persistent and anti-persistent states (Fig.\ \ref{fig7}) using the non-parametric Kaplan–Meier estimator \cite{KM58}. 
Survival analysis revealed that the majority of hemocytes spent longer in persistent than anti-persistent states. The duration of anti-persistent motion increases in embryos from the control to \emph{SCAR} in the order control$<${\it LanB1}$<$\emph{SCAR}. This agrees with the visual inspection of trajectories in Fig. \ref{fig:3dtracks} (a,b,c). The reduction of persistence of the movement of the hemocytes in {\it LanB1} and \emph{SCAR} embryos compared to the  control is also corroborated by the MSDs (Fig.\ref{fig2} (b)). Furthermore, the largest difference of survival times between persistent and anti-persistent states is exhibited by the control embryo hemocytes. The {\it LanB1} mutation reduces the survival times of the persistent states of motion while keeping anti-persistent states the same as the control. In contrast, the \emph{SCAR} mutation increases the survival times of the anti-persistent states of motion while keeping persistent states the same as the control. Thus, {\it LanB1} and \emph{SCAR} mutations produce significant differences in hemocyte migration. This suggests that actin polymerisation is responsible for regulating the anti-persistent states of motion, whereas laminin is responsible for persistent states of motion. This is the first time evidence has been provided that external and internal cellular mechanisms contribute to different aspects of cellular migration in a population of cells \emph{in vivo} in a distinct and quantifiable manner. Finally, this evidence confirms that the incremental displacement steps and incremental angles between steps shown in Figs. \ref{fig2}d and e are connected to specific mutations within the hemocytes. 



\section*{Discussion}
We present the first quantitative analysis of hemocyte migration in three dimensions inside live \emph{Drosophila} embryos and provide strong evidence that hemocyte motility must be considered within the framework of anomalous transport \cite{Bressloff}. The average Hurst exponents for all the cells decreased with time as the embryos aged and the cell motility became both less motile and less persistent, with many cells becoming trapped in localized regions. The control, \emph{LanB1} and \emph{SCAR} embryos had distinct patterns of anomalous motility. The control hemocytes quickly dispersed to perform their diverse biological functions. In \emph{LanB1} the process of dispersal was significantly slowed and the embryos failed to hatch into larvae. In \emph{SCAR} hemocytes dispersion was even more significantly disrupted and the mutation also made the embryos unviable. 

A broad spectrum of Hurst exponents was found for the control, \emph{LanB1} and \emph{SCAR} embryos  (Fig. \ref{fig3}(d)). The distributions of Hurst exponents were characteristic of the type of mutant considered. The hemocytes are switching between different levels of persistent motility rather than there being distinct subpopulations of cells that have constant amounts of persistent motility (four populations of gait were identified by a Gaussian mixture model, SI Fig. \ref{fig:gmmh}).

Detailed analysis of the statistics of motion showed distinct changes with the type of embryo considered: control, \emph{LanB1} and \emph{SCAR}. These could be explained in terms of the known biological functions of the mutants. \emph{LanB1} affected laminin in the extracellular matrix and thus the local viscoelasticity. The \emph{LanB1} mutant (deduced to increase the local viscoelasticity) reduced the amplitude of the motion (observed via the MSDs), decreased the persistence of motion (observed via Hurst exponents and angular distributions), but did not change the gait of the motility (observed in the velocity autocorrelations). \emph{SCAR} directly affected the motility by perturbing intracellular actin polymerisation. This caused the largest reduction in the amplitude of the motion observed (via the MSDs), decreased the persistence of motion (observed via Hurst exponents and angular distributions) and had a significant change to the gait of the motility (observed in the velocity autocorrelations).  

The survival probabilities for the durations of persistent motion followed truncated power law distributions (Fig. \ref{fig7}). The distributions of displacement increments were non-Gaussian with exponential tails (Laplace-like) in common with other previous studies of single cell motility \cite{Wu2014}. In terms of their spatial distribution, hemocytes began densely packed in a small area (the head mesoderm) and rapidly spread around the embryo. In \emph{LanB1} embryos, anti-persistent motion dominates from an earlier stage than with the control and \emph{SCAR}, highlighting the influence of the local viscoelasticity on the trapping of cells in antipersistent states. 

From the perspective of theoretical modelling, the equivalence of the MSD and TAMSD implies the cellular motion is ergodic (Fig.\ \ref{fig2}(b)). This allows CTRW models to be discounted and implies FBM or fractional Langevin equation (FLE) models may be relevant. It also allow glassy dynamics to be discounted, observed previously in jammed embryonic cells \cite{Schotz}, since it would again be non-ergodic. The major difference between the FBM and the FLE models is that the FLE fulfills the fluctuation-dissipation theorem while the FBM does not. Since most cellular and intracellular processes consume energy (e.g. ATP), they break the conditions for thermal equilibrium needed for the fluctuation-dissipation theorem. Therefore, the FBM model seems preferable to describe cell motility.

In general, heterogeneous FBM appears to be a very flexible model to describe anomalous transport in biology provided its assumptions are fulfilled i.e. ergodicity, bounded Hurst exponents in the range $[0,1]$ etc. A previous study showed hFBM could be used to describe endosomal transport in human cells \cite{han2020deciphering, Korabel_etal, Korabel_etal_Entropy}. Here we use it to describe the motion of whole cells on much larger time and length scales. Although both hemocyte and endosomal motility were found to be heterogeneous in time and space, crucial differences are seen in the statistics. Hemocyte displacements are found to be Laplace-like, whereas endosomes follow power law distributions. The reasons for these non-Gaussian distributions of displacements were found to be in the power law distributions of endosome diffusion coefficients \cite{Korabel_etal, Korabel_etal_Entropy} and Weibull-like distributions of diffusion coefficients of hemocytes (see Figs.\ \ref{fig3}). Curiously, the generalised diffusion coefficients for hemocytes are anticorrelated, whereas those for endosomes are correlated. A loose analogy would be to a driver of a vehicle hitting the brake and accelerator pedals at the same time with hemocytes and hemocyte motility must be under careful control by a feedback loop \cite{Davis2015}. Furthermore the motion of neighbouring hemocytes are found to be correlated, whereas endosomes are not. The hFBM model accurately predicts the distributions of displacements given the distribution of generalized diffusion coefficients of the hemocyte cells in embryos from the experiments (Fig.\ \ref{fig4}) and it is consistent with all the other statistical data extracted. This is a major success for the hFBM model in this context.

It is fascinating to observe the synchronisation of the motion of neighbouring, but non-contacting, hemocytes  in embryos. The exact biological origin of this synchronisation would require further research, but chemotaxis in response to a common small molecule signal (e.g. Pvf \cite{Wood2006}) is a primary candidate to explain the motion of non-contacting cells (Fig.\ \ref{fig5}). Upon contact the anomalous transport of hemocytes continues to be synchronised, but the directions become randomized due to cell scattering. Contact inhibition of locomotion (CIL) is hypothesized to make a major contribution to the dispersal of hemocytes during development \cite{Pocha, Davis}. Classically Abercrombie defines two types of CIL \cite{Abercrombie}. In type I CIL cells stop following one another after contact, whereas in type II CIL cells repolarise after contact and migrate away from one another. In contrast, our data indicates a more nuanced model for hemocyte motion upon contact, i.e. anomalous CIL, in which the amplitude, scaling and persistence of the anomalous transport becomes synchronised upon contact (an anomalous tango of hemocytes). This moves beyond simple notions of cells in pure stationary, diffusive or ballistic states. Coherent motion due to large scale embryonic remodelling (e.g. germ band retraction and nerve cord condensation) is also expected to contribute, but there would be no time delays in the oscillations in this case.

Some \emph{in vitro} biophysical work has previously been performed on CIL with human cancer cells in microfluidic geometries \cite{Bruckner} and machine learning techniques were used to classify the processes involved in cell scattering. Here, we extend the biophysical analysis of CIL and provide clear evidence that anomalous cellular transport is important with hemocytes i.e. anomalous CIL occurs, characterised by synchronised fractional kinetics \emph{in vivo}. 

Oscillations in the motion of hemocytes on the time scale of $>$2 mins have been previously described in the biological literature due to oscillations of hemocyte protrusions \cite{Wood2006}. The oscillations of hemocyte protrusions were characterised by increases in the polarity of cells i.e. the contact area of the cells oscillated along the directions of motion. We expect these oscillations in hemocyte polarity also explain the oscillations in anomalous motility and anomalous CIL, although we have not measured the polarity directly (only the nuclei were labelled in our experiments, not the external cellular membrane).

A huge amount of biologically relevant information is available from cell tracking experiments in embryos. Future research on anomalous transport could investigate other cells beyond hemocytes, later stages of morphogenesis and other types of embryos e.g. zebra fish or murine. Heterogeneous fractional Brownian motion informed by neural networks is a powerful new tool to automatically segment the dynamics of cell motility in time and space. Future studies could link the generalized diffusion coefficients and anomalous exponents with specific molecular processes in the cells \cite{Davis2015}. This would provide an accurate quantitative model to connect the molecular biology with the cell motility over the complete duration of the embryonic stage of development. For example, a current challenge is to explain the anti-correlation of the generalized diffusion coefficients and the anomalous exponents in oscillatory hemocyte motility, which has not previously been observed in the field of anomalous transport. The feed forward neural network used provides state of the art dynamic segmentation of the tracks, but it is possible that other architectures could improve on this performance e.g. the Wavenet convolutional neural network with causality constraints and a wider range of dynamic sensitivity \cite{vandenOord}. Lattice light sheet microscopy could allow the tracking of smaller features inside cells with higher resolution \cite{Chen}, although it is expected SPIM will still be competitive for the analysis of whole cell motions.

\section*{Methods}
\subsection*{Drosophila husbandry}
Drosophila stocks were raised  on iberian food according to standard procedures. Flies were maintained in incubators at 25$^{\circ}$C and controlled humidity. For embryo collection, flies were placed in cages and left to lay eggs on apple juice plates prepared in house.
The following Drosophila stocks were used: \emph{SCAR} delta37 \cite{Evans}, \emph{LanB1} def \cite{Sanchez} and \textit{w$^{1118}$}(BL$\#$5905) 
served as wild-type control. Hemocyte nuclei were labelled by expression of Histone2A-mCherry under the direct control of the {\it serpert} promoter \cite{Evans}. Embryos of the appropriated genotype were identified based on the absence of balancer chromosomes carrying fluorescent markers.

\subsection*{Preparation of embryos for imaging and Selective plane illumination microscopy (SPIM)}
Embryos of an overnight collection were dechorionated in 50\% bleach for 1 min and extensively washed in water. Embryos at stage 12 of development (440-580 min AEL) were selected for SPIM imaging and mounted on a cylinder of 1\% low melting point agarose (Sigma Aldrich).

Embryos were imaged using a Carl Zeiss Z1 lightsheet microscope equipped with a 20x/1.0NA Plan Aprochomat objective. 1200 frame movies of the embryos in three dimensions were created. The time interval between consecutive frames was 17.6 seconds. The excitation wavelength was 561 nm. Cells were detected and tracked using the IMARIS software package (version 8.2, Oxford Instruments, United Kingdom).

\subsection*{Neural networks}
A feedforward neural network (FNN) architecture was used to dynamically segment tracks for hemocyte cells that was trained on fractional Brownian motion \cite{han2020deciphering}. The neural network takes the incremental displacements of a trajectory and returns the time-dependent Hurst exponent $0<H(t)<1$ which is a measure of persistence of the motion. For a standard fractional Brownian motion, the Hurst exponent is constant and the mean squared displacement grow as $MSD(t) \simeq t^{2H}$. Our previous FNN study used two dimensional tracks from intracellular endosomes, but the extension to the three dimensional motion of whole cells (specifically the nuclei of whole cells) was straightforward.

\subsection*{Statistical analysis of experimental hemocyte's trajectories}

The ensemble-averaged mean squared displacement (MSD) of 3D experimental hemocyte's trajectories $\{x_i, y_i, z_i\}$ is defined as:
\begin{equation}
\mbox{MSD}(t) = \frac{1}{l^2} \left< (x_i(t)-x_i(0))^2 + (y_i(t)-y_i(0))^2 + (z_i(t)-z_i(0))^2 \right> = 6 D_{\alpha} \left( \frac{t}{\tau} \right)^{\alpha}.
\label{emsd}
\end{equation}
The angled brackets denotes averaging over an ensemble of trajectories, $\left< A \right>=\sum_{i=1}^{N} A_i/N$, where $N$ is the number of trajectories in the ensemble. We set the time scale $\tau=1$ sec and the length scale $l=1$ $\mu$m in order to make the generalized diffusion coefficient $D_{\alpha}$ dimensionless. The MSD is fitted to the power law function to extract the anomalous exponent $\alpha$ and the generalized diffusion coefficient $D_{\alpha}$. Notice that $\alpha$ and $D_{\alpha}$ are constants which characterize averaged transport properties of the ensemble of trajectories. 

The time-averaged mean squared displacement (TMSD) of a individual trajectory of a duration $T$:
\begin{equation}
\mbox{TMSD}_i(t) = \frac{1}{l^2} \frac{1}{T-t} \int_{0}^{T-t} \left( x_i(t'+t)-x_i(t'))^2 + (y_i(t'+t)-y_i(t'))^2 + (z_i(t'+t)-z_i(t'))^2 \right) dt', 
\label{tmsd}
\end{equation}
where $l$ is the length scale which we choose to be $l=1$ $\mu$m. 
TMSDs of individual trajectories can be averaged further over the ensemble of trajectories: 
\begin{equation}
\mbox{TMSD}(t) = \left< \mbox{TMSD}_i(t) \right>.
\label{etmsd}
\end{equation}
We calculated the local TMSD (L-TMSD) e.g. TMSD$_i$ in a time window $(t-W/2,t+W/2)$ which is shifted along a trajectory and L-TMSD is calculated in each window. By fitting the first $10$ L-TMSDs points to power law functions, we extracted the time dependent local anomalous exponents $\alpha_L(t)$ and the time dependent local generalized diffusion coefficients $D_{\alpha_L}(t)$:
\begin{equation}
\mbox{L-TMSD}(t) = 6 D_{\alpha_L} \left( \frac{t}{\tau} \right)^{\alpha_L}.
\label{ltmsdfit}
\end{equation}
Again, the time scale $\tau=1$ sec and the length scale $l=1$ $\mu$m are introduced in order to make the local generalized diffusion coefficients dimensionless and facilitate comparison. 

The time dependence of the MSD determine the character of the movement of the ensemble of hemocytes as normal diffusion for $\alpha = 1$, sub-diffusion for $\alpha < 1$ and super-diffusion for $\alpha > 1$. Similarly the time dependence of the L-TMSD determine the character of local movement of a single hemocyte trajectory.

\subsection*{Velocity auto-correlation function of experimental trajectories}

The velocity auto-correlation function (VACF) is defined as: 
\begin{equation}
\mbox{VACF}(t) = \frac{ \int_{0}^{T-t - \delta} \vec{v}(t'+t) \vec{v}(t') dt'}{T-t - \delta}.
\label{tvacf}
\end{equation}
where $\vec{v}=\frac{\vec{r}(t+\delta)-\vec{r}(t)}{\delta}$ and $\delta=17.6$ s. To improve statistics, the VACF is further averaged over the ensemble of all the trajectories.
 
\subsection*{Heterogeneous Fractional Brownian Motion}
Heterogeneous FBM trajectories were simulated using the overdamped Langevin equation with generalized diffusion coefficient $D$ and Hurst exponent $H$:
\begin{equation}
\frac{d\mbox{r}(t)}{dt} = \sqrt{2 D} \mbox{f}^{fGn},
\label{LE}
\end{equation}
where $\mbox{f}^{fGn}=\{f^{fGn}_x,f^{fGn}_y,f^{fGn}_z\}$ is the vector of fractional Gaussian noise with correlation function 
\begin{equation}
\left< \mbox{f}^{fGn}(t_1) \mbox{f}^{fGn}(t_2) \right> \simeq H (2H-1) |t_1 - t_2|^{2H-2}.
\label{fGn}
\end{equation}
We generated $\mbox{f}^{fGn}$ in MATLAB (The MathWorks, Nantick, MA) using ffgn function.  In the simulations shown in Fig.\ \ref{fig4}, the Hurst exponents were constant ($H=0.65$) for all trajectories. The generalized diffusion coefficients were assigned from the Weibull distribution  $f(D) = c (D^{c-1}/D_0^{c}) \exp(-(D/D_0)^c)$ with $c=0.33$ and $D_0=0.0018$ which were determined from haemocyte trajectories (see Fig.\ \ref{fig3}c). The PDF of $D$ was truncated at $D_{min}=0.1$ and $D_{max}=100$.

\subsection*{Survival analysis} 
We used instantaneous Hurst exponents $H(t)$ to segment haemocyte trajectories into persistent ($H(t)>0.5$) and anti-persistent ($H(t)<0.5$) states of motion. The survival time in a persistent or anti-persistent states is than defined as a time the cell remains in the same state for time $t$. Kaplan-Meier non-parametric estimator of the time dependent survival probability that an event has not occurred at a time $t_i$ is defined as $\Psi(t_i) = \left(1- \frac{d_i}{n_i} \right) \Psi(t_{i-1})$, where $d_i$ is the number of events that occur at time $t_i$ and $n_i$ is the number of events survived up to time $t_{i}$. By definition, $\Psi(t_0)=1$.

\section*{Acknowledgements}
DH acknowledges financial support from the Wellcome Trust Grant No. 215189/Z/19/Z. NK acknowledges financial support from EPSRC Grant No. EP/V008641/1. TM acknowledges financial support from the Wellcome Trust Grant no. 201958/Z/16/Z. We thanks the MPhys students for their preliminary contributions: Tom Hughes-Jordan, Jack Forrester, Nicole Rolph and Charlotte Payne. Viki Allan and Sergei Fedotov are thanked for many useful conversations. We thank the University of Manchester Fly and Bioimaging Facilities, in particular David Spiller, for assistance with microscopy and image analysis.

\section*{Author contributions statement}

\section*{Competing financial interests} 
The authors declare no competing interests.

\section*{Data availability}
The data that support the findings of this study are available from the corresponding author upon reasonable request.


\begin{thebibliography}{10}

\bibitem{Metzner} Metzner C., Mark, C. Steinwachs, J. Lautscham, L. Stadler F. and Fabry B., Superstatistical analysis and modelling of heterogeneous random walks. Nat. Commun. {\bf 6}, 7516 (2015).

\bibitem{Selmeczi} Selmeczi D., Mosler S., Hagedorn P.H., Larsen N.B. and Flyvbjerg H., Cell motility as persistent random motion: theories from experiments, Biophys. J. {\bf 89}, 912–931, (2005).

\bibitem{Wu2014} Wu P.-H., Giri A., Sun S. X. and Wirtz D., Three-dimensional cell migration does not follow a random walk, Proceedings of the National Academy of Sciences {\bf 111}, 3949-3954 (2014). 

\bibitem{Wu2015} Wu P.-H., Giri A.. and Wirtz D. Statistical analysis of cell migration in 3D using the anisotropic persistent random walk model. Nat Protoc {\bf 10}, 517–527 (2015).

\bibitem{Huda} Huda S., Weigelin B., Wolf K., Tretiakov K.V., Polev K., Wilk G., Iwasa M., Emami F.S., Narojczyk J.W., Banaszak M., Soh S., Pilans D., Vahid A., Makurath M., Friedl, P., Borisy G.G., Kandere-Grzybowska K. and Grzybowski B.A. Nat. Commun. {\bf 9}, 4539
(2018).

\bibitem{Harris} Harris T.H., Banigan E.J., Christian D.A., Konradt C., Tait Wojno E.D., Norose K., Wilson E.H., John B., Weninger W., Luster A.D., Liu A.J. and Hunter, C.A., Generalized Lévy walks and the role of chemokines in migration of effector CD8(+) T cells. Nature {\bf 486}, 545–548 (2012).


\bibitem{Ariel} Ariel G., Rabani A., Benisty S., Partridge J.D., Harshey R.M. and Be'er, A. Swarming bacteria migrate by Lévy walk, Nat. Commun. {\bf 6}, 8396 (2015).

\bibitem{WirtzReview} Wu P.-H., Gilkes D.W. and Wirtz D.
Annual Review of Biophysics {\bf 47}, 549-567 (2018).

\bibitem{Tomer} Tomer R., Khairy K., Amat F. and Keller P. J., Quantitative high-speed imaging of entire developing embryos with simultaneous multiview light-sheet microscopy.
Nature Methods {\bf 9}, 755–763 (2012). 

\bibitem{Schott} Schott B., Traub M., Schlagenhauf C., Takamiya M., Antritter T., {\it et al.} EmbryoMiner: A new framework for interactive knowledge discovery in large-scale cell tracking data of developing embryos. PLOS Computational Biology {\bf 14}, e1006128 (2018).

\bibitem{Chen} Chen B. C. {\it et al.} Lattice light-sheet microscopy: imaging molecules to embryos at high spatiotemporal resolution. Science {\bf 346}, 1257998 (2014).

\bibitem{Kim} Kim S., Pochitaloff M., Stooke-Vaughan G.A. {\it et al.} Embryonic tissues as active foams. Nat. Phys. {\bf 17}, 859-866 (2021). 

\bibitem{Schotz} Sch\"{o}tz E-M., Lanio M., Talbot J.A., Manning, M.L. Glassy dynamics in three-dimensional embryonic tissues. J. R. Soc. Interface {\bf 10}, 20130726 (2013).

\bibitem{Goodwin} Goodwin, K. and Nelson, C. M. Mechanics of development. Dev. Cell {\bf 56}, 240-250 (2021).

\bibitem{Huisken} Huisken J. {\it et al.} Optical sectioning deep inside live embryos by selective plane illumination microscopy. Science {\bf 305}, 1007–1009 (2004).

\bibitem{Korabel_etal} Korabel N., Han D., Talonni A., Pagnini G., Fedotov S., Allan V.J. and Waigh T.A., arXiv:2107.07760 (2021).

\bibitem{Korabel_etal_Entropy} Korabel N., Han D., Talonni A., Pagnini G., Fedotov S., Allan V.J. and Waigh T.A., Local Analysis of Heterogeneous Intracellular Transport:
Slow and Fast Moving Endosomes. Entropy {\bf 23}, 958 (2021).

\bibitem{han2020deciphering} Han D., Korabel N., Chen R., Johnston M., Gavrilova A., Allan V.J., Fedotov S. and Waigh T.A. Deciphering anomalous heterogeneous intracellular transport with neural networks. Elife {\bf 9}, e52224 (2020).


\bibitem{Caspi} Caspi A., Granek R., and Elbaum M. Enhanced diffusion in active intracellular transport. Phys.\ Rev.\ Lett.\ {\bf 85}, 5655 (2000).

\bibitem{Granick1} Wang B., Anthony S.\ M., Bae S.\ C. and Granick, S. Anomalous yet Brownian. PNAS {\bf 106}, 15160 (2009). 

\bibitem{Granick2} Wang B., Kuo J., Bae S.\ C., Granick S.\ When Brownian diffusion is not Gaussian. Nat. Mater. {\bf 11}, 481-485 (2012).

\bibitem{PLOS} Korabel N., Waigh T.A., Fedotov S., Allan V.J. Non-Markovian intracellular transport with sub-diffusion and run-length dependent detachment rate. PLoS ONE {\bf 13}, e0207436 (2018). 

\bibitem{Fedotov} Fedotov S., Korabel N., Waigh T.A., Han D., Allan V.J. Memory effects and L\'evy walk dynamics in intracellular transport of cargoes. Phys. Rev. E {\bf 98}, 042136 (2018).

\bibitem{Kenwright} Kenwright D.A., Harrison A.W., Waigh T.A., Woodman P.G., Allan V.J. First-passage-probability analysis of active transport in live cells. Phys Rev E {\bf 86}, 031910 (2012).

\bibitem{Cherstvy} Cherstvy A.G., Nagel O., Beta C. and  Metzler R., Non-Gaussianity, population heterogeneity, and transient superdiffusion in the spreading dynamics of amoeboid cells, Phys. Chem. Chem. Phys. {\bf 20}, 23034-23054 (2018).

\bibitem{Takagi} Takagi H., Sato M.J., Yanagida T. and Ueda M., Functional Analysis of Spontaneous Cell Movement under Different Physiological Conditions. PLoS ONE {\bf 3}, e2648 (2008). 

\bibitem{Klages} Dieterich P., Klages R., Preuss R. and Schwab A. Anomalous dynamics of cell migration. Proceedings of the National Academy of Sciences {\bf 105}, 459-463 (2008).

\bibitem{Alert} Alert R. and Trepat X., Physical models of collective cell migration, Annu. Rev. Condens. Matter Phys. {\bf 11}, 77-101 (2020).

\bibitem{Romanczuk} Romanczuk, P., Bär, M., Ebeling, W. {\it et al.} Active Brownian particles. Eur. Phys. J. Spec. Top. {\bf 202}, 1–162 (2012).

\bibitem{Wood} Wood W. and Jacinto A., Drosophila melanogaster embryonic haemocytes: masters of multitasking. Nat. Rev. Mol. Cell Biol. {\bf 8}, 542-551 (2007).

\bibitem{Davis} Davis J. R., Huang C.-H., Zanet J., Harrison S., Rosten E., Cox S., Soong D. Y., Dunn G. A., and Stramer B. M. Emergence of embryonic pattern through contact inhibition of locomotion, Development {\bf 139}, 4555-4560 (2012).

\bibitem{Pocha} Pocha S. M. and Montell D.J. Cellular and Molecular Mechanisms of Single and Collective Cell Migrations in Drosophila: Themes and Variations. Annu. Rev. Genet. {\bf 48}, 295-318 (2014).

\bibitem{Sanchez} Sánchez-Sánchez, B. J., Urbano, J. M., Comber, K., Dragu, A., Wood, W., Stramer, B. and Martín-Bermudo, M. D. {\it Drosophila} Embryonic Hemocytes Produce Laminins to Strengthen Migratory Response.  
Cell Rep. {\bf 21}, 1461-1470 (2017).

\bibitem{Evans} Evans, I. R., Ghai, P. A., Urbančič, V., Tan, K-L. and Wood, W. SCAR/WAVE-mediated processing of engulfed apoptotic corpses is essential for effective macrophage migration in {\it Drosophila}. Cell Death Differ. {\bf 20}, 709-720 (2013).

\bibitem{Metzler2014} Metzler R., Jeon J.-H., Cherstvy A. G., and Barkai E. Anomalous diffusion models and their properties: non-stationarity, non-ergodicity, and ageing at the centenary of single particle tracking. Phys. Chem. Chem. Phys. {\bf 16}, 24128 (2014).

\bibitem{Stramer} Stramer B. and Mayor R.  Mechanisms and in vivo functions of contact inhibition of locomotion. Nat. Rev. Mol. Cell Biol. {\bf 18}, 43 (2017).

\bibitem{KM58} Kaplan, E. L., and Meier, P. Nonparametric estimation from incomplete observations, J. Am. Stat. Assoc. {\bf 53}, 457-481 (1958).

\bibitem{Bressloff} Bressloff P. C., Stochastic Processes in Cell Biology(Springer, New York, 2014).

\bibitem{Davis2015} Davis, J. R. et al, Cell {\bf 161}, 361-373 (2015).

\bibitem{Wood2006} Wood W., Faria C. and Jacinto A. Distinct mechanisms regulate hemocyte chemotaxis during development and wound healing in Drosophila melanogaster. J. Cell Biol {\bf  173}, 405-416 (2006).

\bibitem{Abercrombie} Abercrombie M. and Heaysman J. E. Observations on the social behaviour of cells in tissue culture, Exp. Cell Res. {\bf 6}, 293-306 (1954). 

\bibitem{Bruckner} Br\"uckner D. P., Arlt N., Fink A., Ronceray P.,  Rädler J. O., and Broedersz C. P. Learning the dynamics of cell-cell interactions in confined cell migration. Proc. Natl. Acad. Sci. U.S.A. {\bf 118},  e2016602118 (2021).


\bibitem{vandenOord} van den Oord A., Dieleman S. {\it et al.}, WaveNet: A Generative Model for Raw Audio. ArXiv (2016).


\end{thebibliography}

\bibstyle{naturemag}

\newpage

\setcounter{figure}{0}
\makeatletter 
\renewcommand{\thefigure}{S\@arabic\c@figure}
\makeatother

\section*{Supplementary Material}


\begin{figure}
    \centering
    \includegraphics[width=\linewidth]{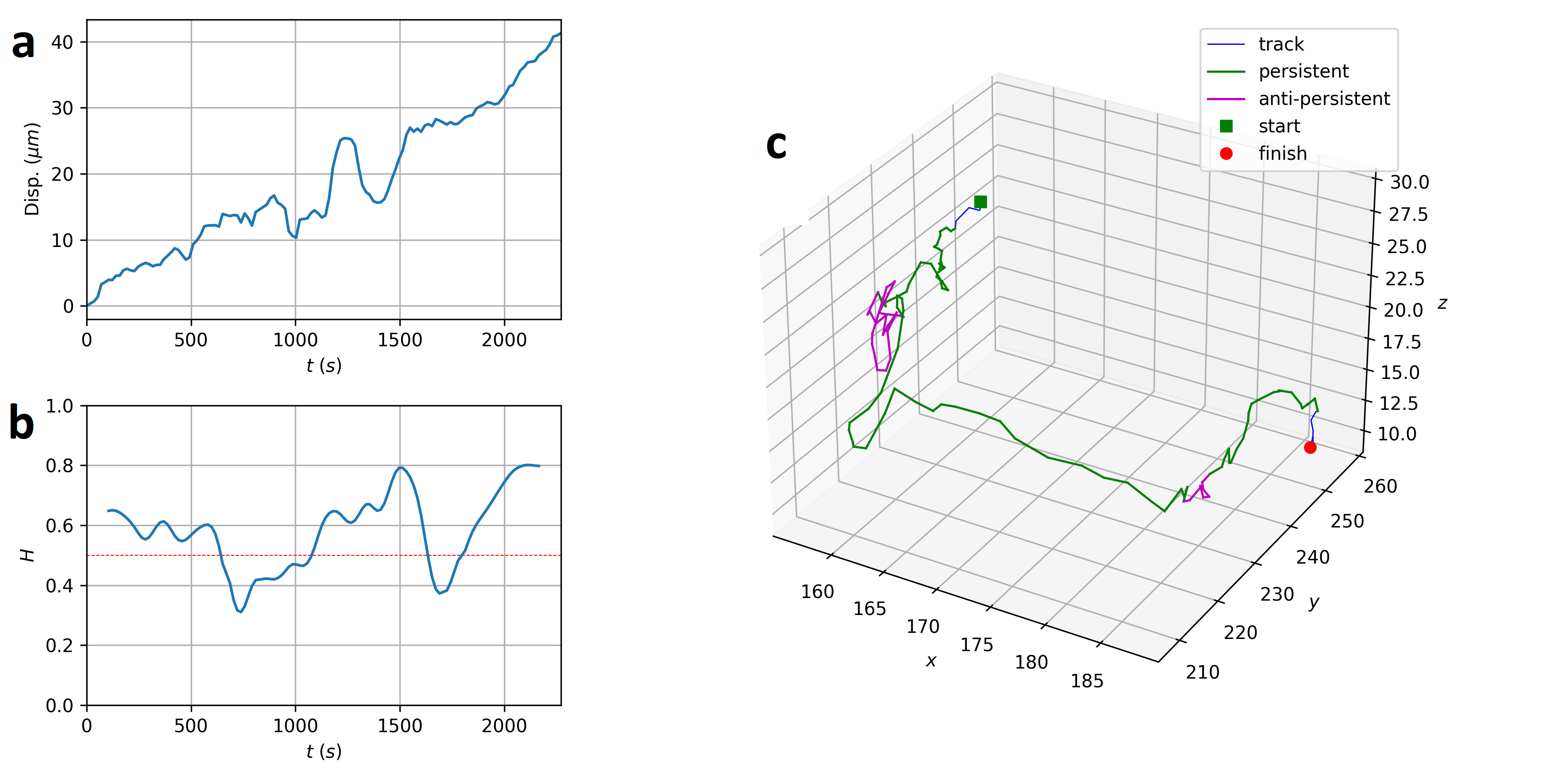}
    \caption{{\bf An example of estimation of instantaneous Hurst exponent $H(t)$ using the neutral network.} See the main text for details. {\bf (a)} shows the displacement and estimated instantaneous Hurst exponent $H(t)$ {\bf (b)} of a sample experimental trajectory {\bf (c)}. Green color in (c) corresponds to persistent motion $H(t)>0.5$ and purple to anti-persistent motion $H(t)<0.5$.}
    \label{fig:trajectory}
\end{figure}

\begin{figure}
    \centering
    \includegraphics[width=\linewidth]{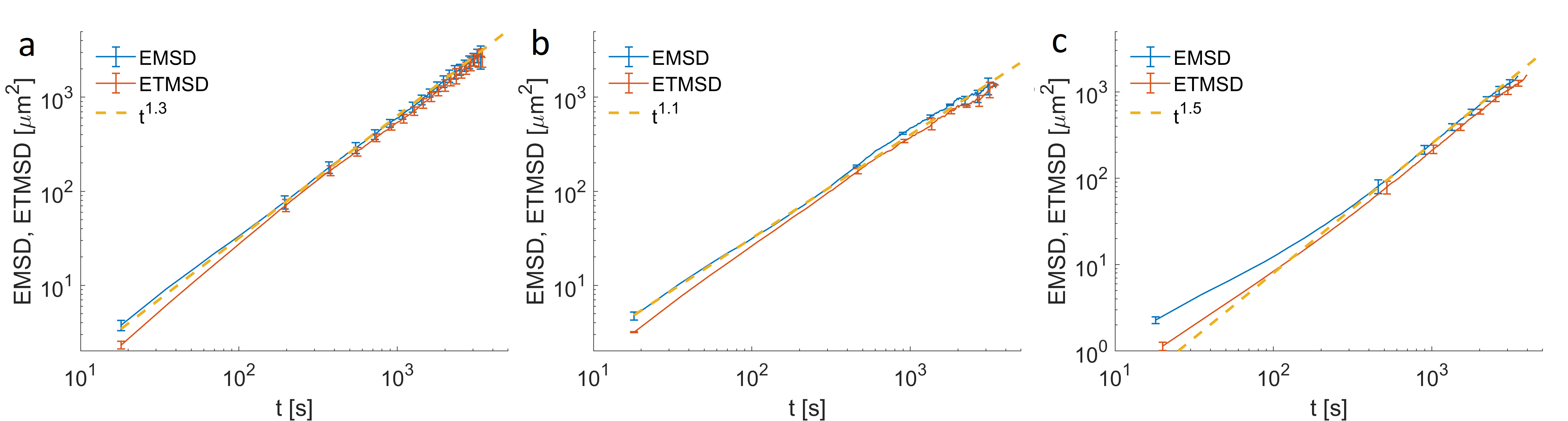}
    \caption{{\bf Ergodic behavior of hemocytes.} The EMSDs and ETMSDS for {\bf (a)} control, {\bf (b)} {\it LanB1} and {\bf (c)} {\it SCAR}  embryos have similar behaviour as functions of time. The dashed line are fits to power laws with exponents $\alpha=1.3 \pm 0.1$ for control, $\alpha=1.1 \pm 0.1$ for {\it LanB1} and $\alpha=1.5 \pm 0.1$ for {\it SCAR} embryos.}
    \label{figS1}
\end{figure}

\begin{figure}
    \centering
    \includegraphics[width=0.7\textwidth]{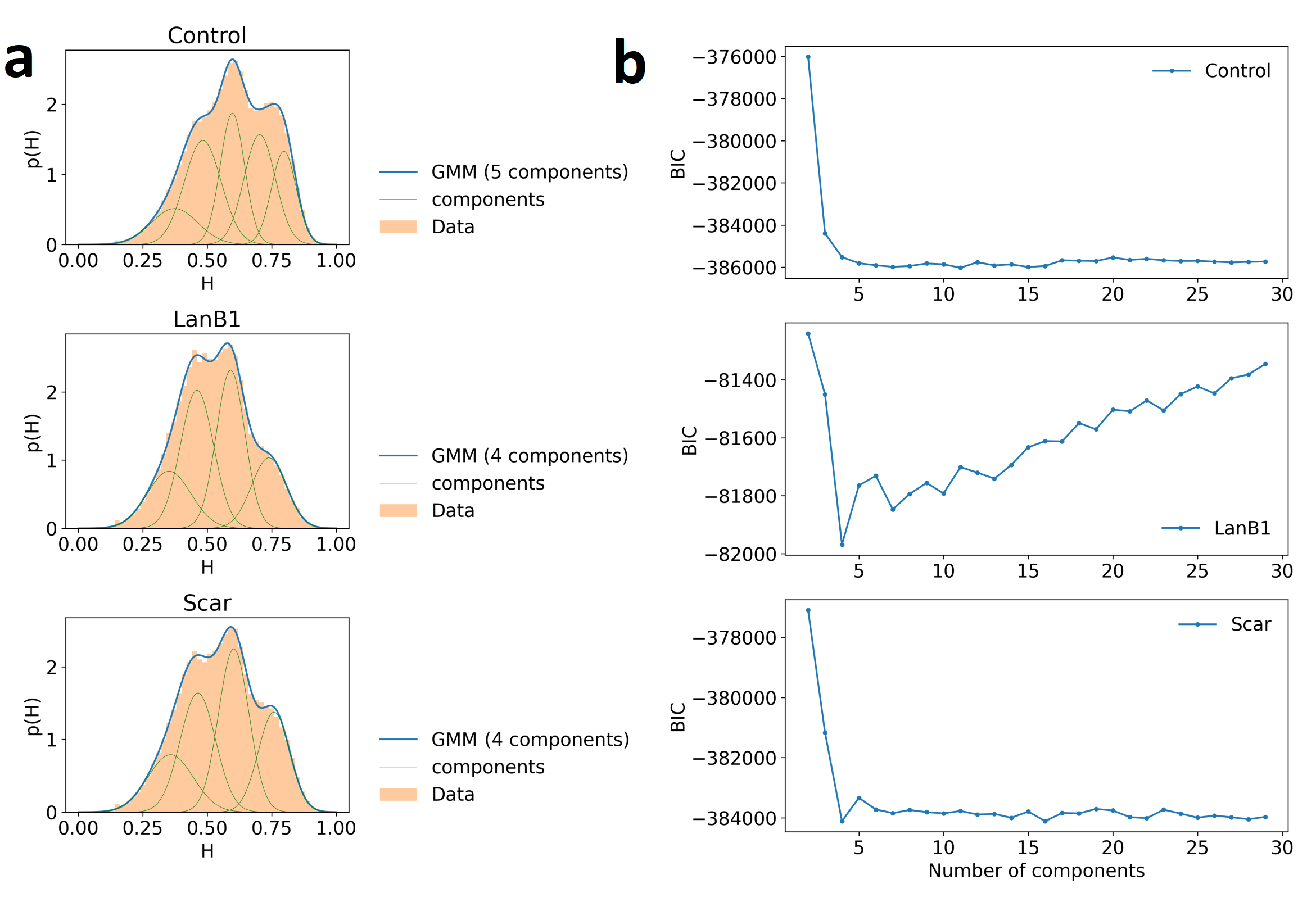}
    \caption{{\bf Fitting the distributions of local Hurst exponents $H(t)$ with Gaussian mixture models}. {\bf (a)} Four separate Gaussian modes were chosen to describe the data which was determined from the BIC criteria {\bf (b)}. The individual Gaussian components in the mixture models show a significant difference in modes of persistent ($H(t)>0.5$) motion between control, {\it SCAR} and {\it LanB1} embryos. The Hurst exponents of modes 1, 2 and 4 decrease with time, whereas 3 is relatively constant and is thought to be associated with passive cell motility due to germ band retraction.
 }
    \label{fig:gmmh}
\end{figure}

\begin{figure}
    \centering
    \includegraphics[width=0.5\textwidth]{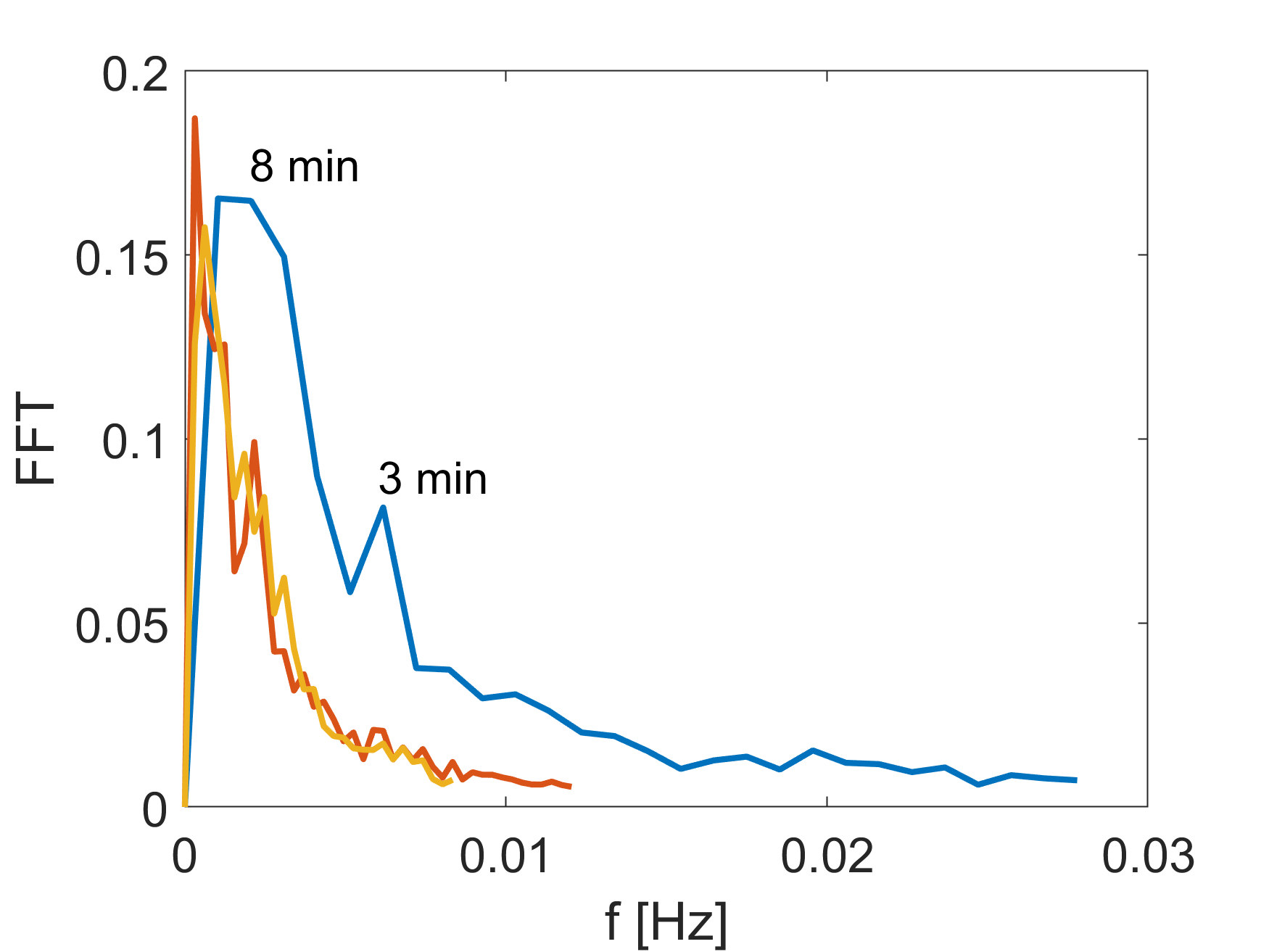}
    \caption{{\bf Fast Fourier transforms (FFT) of local anomalous exponents in control, {\it LanB1} and {\it SCAR} embryos.} FFT of $\alpha_L(t)$ as a function of frequency (f, averaged over $10$ trajectories) for control embryo (blue curve) reveals two oscillation periods of $3$ and $8$ minutes. Period $3$ minutes might be related to the oscillatory behaviour of hemocyte polarization during the motion. FFT of $\alpha_L(t)$ for {\it LanB1} and {\it SCAR} embryos have remnants of the period $3$ minutes and show periods of $8$ and $14$ minutes instead.}
    \label{figFFT}
\end{figure}

\begin{figure}
    \centering
    \includegraphics[width=0.7\textwidth]{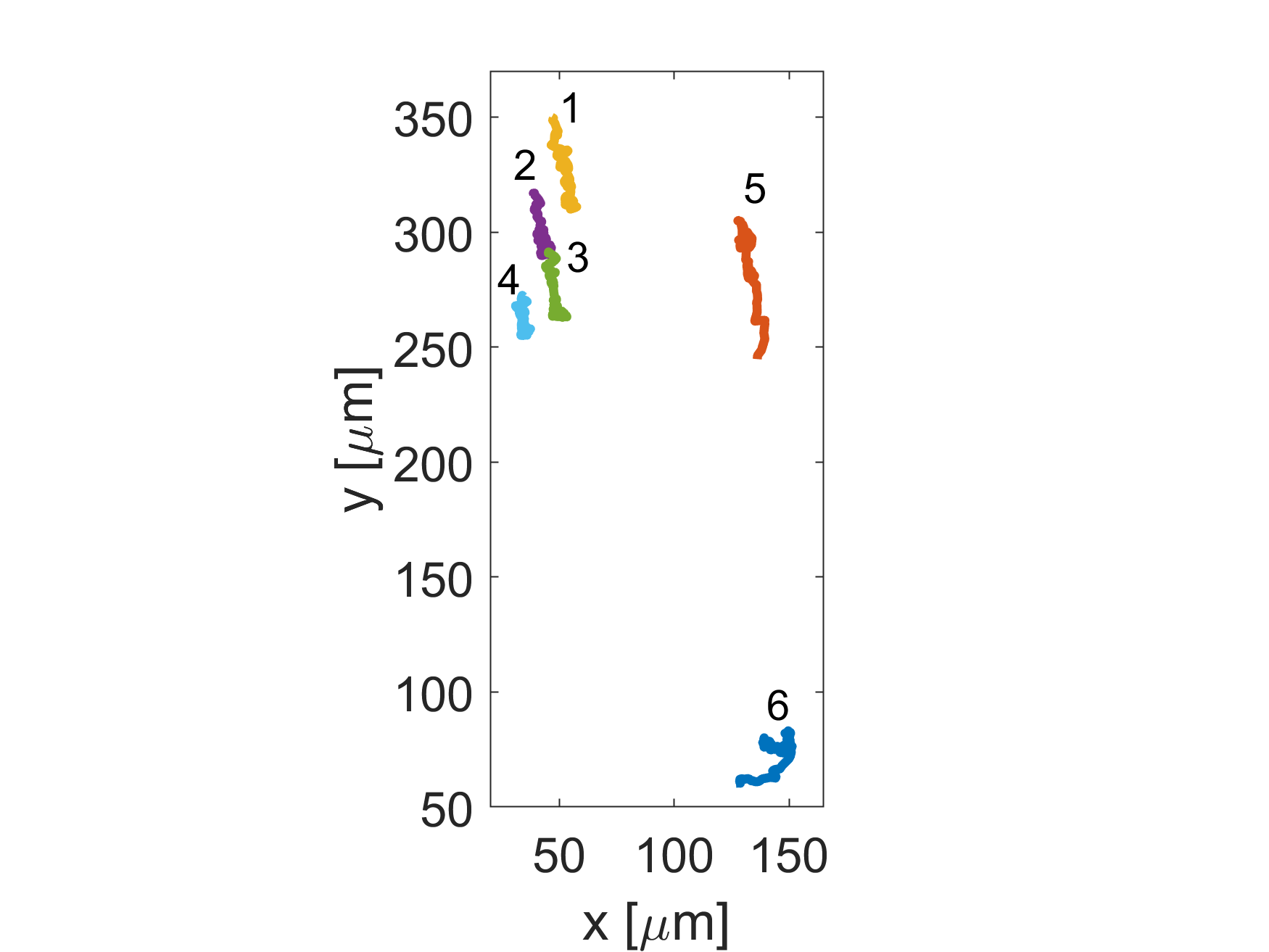}
    \caption{{\bf Quantification of correlated movement of hemocytes in {\it LanB1} embryo.} Correlation coefficients for six sample trajectories shown in the figure is calculated by Corr$_{1,2}=$Corr$(\lVert \bf{r_1} \rVert, \lVert \bf{r_2} \rVert)$. Here $\lVert \bf{r_1} \rVert$ and $\lVert \bf{r_1} \rVert$ are distances vectors of trajectories $1$ and $2$ to the origin as functions of time. The correlation coefficients of pairs of trajectories are Corr$_{1,2}=0.99$, Corr$_{1,3}=0.98$, Corr$_{1,4}=0.97$, Corr$_{1,5}=0.83$ and Corr$_{1,6}=-0.79$. Trajectories $1, 2, 3, 4$ display strongly correlated movement while trajectory $6$ is moving anti-correlated relative to them.}
    \label{fig:oclAlf}
\end{figure}

\begin{figure}
    \centering
    \includegraphics[width=1\textwidth]{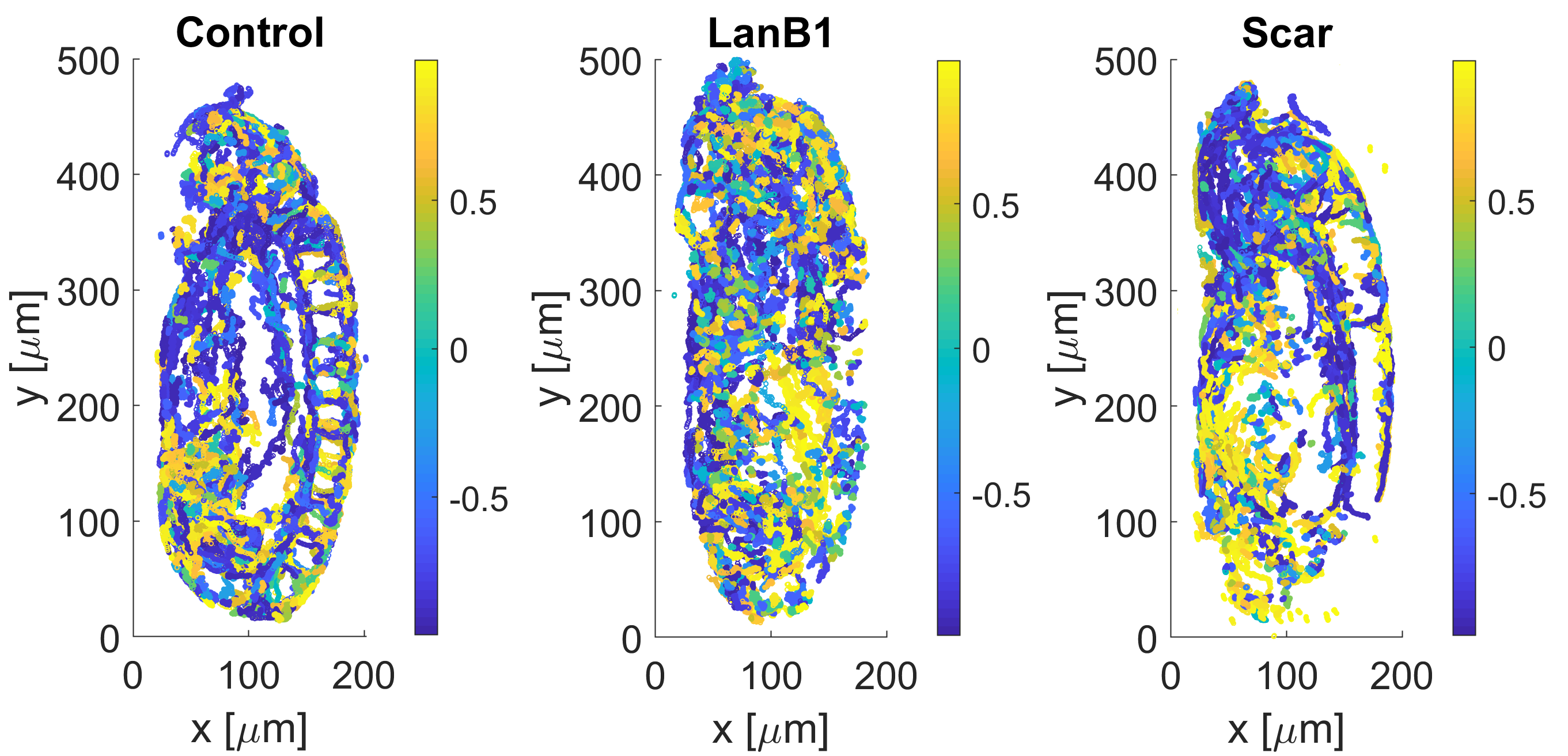}
    \caption{{\bf Quantification of correlated movement of hemocytes in control, {\it LanB1} and SCAR embryos.} Colours of trajectories represent correlation coefficients calculated with respect to the first trajectory.}
    \label{fig:oclAlf1}
\end{figure}

\end{document}